\documentclass[letterpaper, 10 pt, conference]{ieeeconf}  % Comment this line out if you need a4paper

\IEEEoverridecommandlockouts                              % This command is only needed if 
\usepackage{amsmath}
\usepackage{bm}
\usepackage{booktabs}
\usepackage[usenames,dvipsnames]{color}
\usepackage{graphicx}
\usepackage{comment}
\usepackage{tikz}
\usepackage{varwidth}
\usetikzlibrary{calc,intersections,patterns,backgrounds,positioning}

\newlength{\hatchspread}
\newlength{\hatchthickness}
\newlength{\hatchshift}
\newcommand{\hatchcolor}{}
\tikzset{hatchspread/.code={\setlength{\hatchspread}{#1}},
         hatchthickness/.code={\setlength{\hatchthickness}{#1}},
         hatchshift/.code={\setlength{\hatchshift}{#1}},% must be >= 0
         hatchcolor/.code={\renewcommand{\hatchcolor}{#1}}}
\tikzset{hatchspread=3pt,
         hatchthickness=0.4pt,
         hatchshift=0pt,% must be >= 0
         hatchcolor=black}
\pgfdeclarepatternformonly[\hatchspread,\hatchthickness,\hatchshift,\hatchcolor]% variables
   {custom north east lines}% name
   {\pgfqpoint{\dimexpr-2\hatchthickness}{\dimexpr-2\hatchthickness}}% lower left corner
   {\pgfqpoint{\dimexpr\hatchspread+2\hatchthickness}{\dimexpr\hatchspread+2\hatchthickness}}% upper right corner
   {\pgfqpoint{\dimexpr\hatchspread}{\dimexpr\hatchspread}}% tile size
   {% shape description
    \pgfsetlinewidth{\hatchthickness}
    \pgfpathmoveto{\pgfqpoint{\dimexpr\hatchshift-0.15pt}{-0.15pt}}
    \pgfpathlineto{\pgfqpoint{\dimexpr\hatchspread+0.15pt}{\dimexpr\hatchspread-\hatchshift+0.15pt}}
    \ifdim \hatchshift > 0pt
      \pgfpathmoveto{\pgfqpoint{-0.15pt}{\dimexpr\hatchspread-\hatchshift-0.15pt}}
      \pgfpathlineto{\pgfqpoint{\dimexpr\hatchshift+0.15pt}{\dimexpr\hatchspread+0.15pt}}
    \fi
    \pgfsetstrokecolor{\hatchcolor}
    \pgfusepath{stroke}
   }

\definecolor{pedestriangreen}{rgb}{0.0, 0.42, 0.24}
\newtheorem{assumption}{Assumption}

\title{\LARGE \bf Multistage Stochastic Model Predictive Control\\ for Urban Automated Driving}

\author{Tommaso Benciolini, Tim Br\"{u}digam, Marion Leibold% <-this % stops a space
\thanks{T. Benciolini, T. Br\"{u}digam, and M. Leibold are with the Chair of Automatic Control Engineering at the Technical University of Munich, Munich, Germany (email: {\tt\small\{t.benciolini; tim.bruedigam; marion.leibold\}@tum.de}).}%
}

\begin{document}
\pagenumbering{gobble}

\maketitle
\thispagestyle{plain}
\pagestyle{plain}

%%%%%%%%%%%%%%%%%%%%%%%%%%%%%%%%%%%%%%%%%%%%%%%%%%%%%%%%%%%%%%%%%%%%%%%%%%%%%%%%
\begin{abstract}
Trajectory planning in urban automated driving is challenging because of the high uncertainty resulting from the unknown future motion of other traffic participants. Robust approaches guarantee safety, but tend to result in overly conservative motion planning. Hence, we propose to use Stochastic Model Predictive Control for vehicle control in urban driving, allowing to efficiently plan the vehicle trajectory, while maintaining the risk probability sufficiently low. For motion optimization, we propose to use a two-stage hierarchical structure that plans the trajectory and the maneuver separately. A high-level layer takes advantage of a long prediction horizon and of an abstract model to plan the optimal maneuver, and a lower level is in charge of executing the selected maneuver by properly planning the vehicle’s trajectory. Numerical simulations are included, showing the potential of our proposal.
\end{abstract}

%%%%%%%%%%%%%%%%%%%%%%%%%%%%%%%%%%%%%%%%%%%%%%%%%%%%%%%%%%%%%%%%%%%%%%%%%%%%%%%%
\section{INTRODUCTION}

\vspace{-11.5cm}
%\mbox
\parbox[t]{20cm}{\small This~work~has~been~accepted~to~the~IEEE~2021~International~Conference~on~Intelligent~Transportation~Systems.\\
The published version may be found at https://doi.org/10.1109/ITSC48978.2021.9564572.}
\vspace{10.7cm}

Compared to automated highway driving, autonomous driving in urban scenarios presents several new challenges. The urban framework is populated by many traffic participants like cars, cyclists, and pedestrians, showing extremely different behaviours, and generating large uncertainty. Complexity arises also due to the fact that various different maneuvers are possible, depending on the traffic conditions and rules, and on the road geometry. Furthermore, in contrast to the highway framework, maneuver switches appear more frequently, to adapt the driving to the changing environment.

In order to deal with the challenges of the urban environment, learning-based methods are appealing~\cite{ChenEtAl2019b}. They allow for a flexible motion, able to handle several possible configurations with no need for rigidly defined switching rules, which is extremely convenient to face complex situations typical for the European urban road environment~\cite{HawkeEtAl2020}. Yet, safety is hardly provable, requiring additional structures~\cite{ChenEtAl2019}, which increase the complexity of those algorithms.

Alternatively, Stochastic Model Predictive Control (SMPC) represents an attractive option to cope with the urban driving complexity. Model Predictive Control (MPC) computes a sequence of inputs minimizing a cost function over a finite prediction horizon, guaranteeing the satisfaction of a set of constraints. Only the first input of the sequence is applied, and, as soon as a new state measurement is available, the optimal control problem is solved again in a receding horizon manner. The possibility to reward control goals (through the design of the cost function) and to express necessary requirements (constraints) makes MPC suitable for the autonomous driving application~\cite{LevinsonEtAl2011}. To cope with uncertainties, Robust MPC (RMPC) considers a worst-case scenario, ensuring a safe motion for each possible circumstance. In~\cite{SchildbachEtAl2016} safe corridors are derived, where the autonomous vehicle can drive ensuring collision avoidance for any realization of the other vehicles' motion. Still, this method is likely to result in an excessively cautious controller producing unnaturally  conservative behaviors, up to the point that under particular conditions the automated vehicle can not take action at all, as briefly discussed in~\cite{BruedigamEtAl2021}.

Differently from RMPC, in SMPC safety constraints are designed in a probabilistic fashion, and the resulting control action must guarantee that the probability of constraint violation is lower than a user-defined risk parameter. This formulation makes SMPC particularly well suited to deal with the uncertain urban framework, since the choice of the acceptable risk level allows to balance the trade-off between safety and efficiency. SMPC is widely used for autonomous driving in highway-like scenarios~\cite{NguyenEtAl2017,CarvalhoEtAl2014,SuhEtAl2018}, but little attention is dedicated to the typical urban framework, where stops are required to avoid collisions with crossing pedestrians or upcoming vehicles at intersections.

Moreover, it is beneficial to take a long prediction horizon into account, allowing to reveal superior maneuver choices, which would prove to be efficient in the long run. Yet, a detailed model is not necessarily useful to optimally plan long-term maneuvers, and a simplified one, considering the dominant dynamics only, may suffice. Additionally, due to the accumulated action of uncertainty, it is not reasonable to rely on detailed prediction models for long-term predictions. In~\cite{BaethgeEtAl2016,BruedigamEtAl2020b} a detailed model is used for short-term predictions, and a simplified one for long-term predictions. Furthermore, a long prediction horizon causes a computational burden difficult to cope with in a quick real-time computation. Move blocking approaches~\cite{ShekharEtAl2015} and non-uniformly spaced horizon techniques~\cite{El-FekyEtAl2017} are proposed to tackle the computational burden resulting from a long horizon. However, these methods do not allow to explicitly distinguish between short-term and long-term actions, solving both at once. This is not ideal for autonomous urban driving, where we can differentiate between trajectory planning, concerning low-level control actions (instant acceleration, steering angle) to drive the vehicle in the short term, and maneuver planning, focusing on more high-level actions (cruise speed planning, lane changing) to be determined considering a longer horizon. It is convenient to deal with maneuver planning and trajectory planning separately, since they have different time scales, and need to be updated with different frequency.

In order to separate the maneuver and trajectory planning, a hierarchical MPC approach is well suited. This technique is extensively employed to handle systems with multiple time scales~\cite{BrdysEtAl2008}. The usage of hierarchical structures in the framework of autonomous driving has already been proposed~\cite{DinhEtAl2020},~\cite{ZieglerEtAl2014}. In~\cite{WangEtAl2015} a high-level layer is used to select discrete maneuver states through pre-defined switching rules, resulting in different setups (e.g., tuning the weighting matrices in the objective function) for the underlying model predictive trajectory guidance module. However, these works take advantage of the high level of the hierarchy to decide on the maneuver planning in the short term, whereas in urban automated driving it is beneficial to use the high-level reasoning for a long-horizon optimization.

The main contribution of this work is applying SMPC to autonomous urban driving. Starting from a stochastic description of the future motion of the other traffic participants (other vehicles, pedestrians), safety conditions for the ego vehicle are derived, resulting in constraints ensuring that the probability of collision remains below the user-defined acceptable risk level. Moreover, to properly plan the motion in the long run, we propose to use a hierarchical structure composed of two layers. On the one hand, the high-level maneuver planner uses a simple model and a large sampling time, and takes advantage of a long horizon to optimize the maneuver. On the other hand, the low-level trajectory planner is in charge of determining the short-term optimal control actions to execute the selected maneuver, considering the detailed model. This control structure allows to benefit from a long-term view, without increasing the computational effort of the trajectory planner. Moreover, maneuver planning and trajectory planning are solved independently, explicitly optimizing each task separately, taking into account different characteristic time scales.

The reminder of the paper is organized as follows. In Section~\ref{sec:overview} an overview of the control hierarchy is outlined. Section~\ref{sec:dynamicalmodels} presents the dynamical models used. The details concerning the low-level trajectory planner and the high-level maneuver planner are given in Section~\ref{sec:trajectoryplanning} and in Section~\ref{sec:maneuverplanning}, respectively. Numerical simulations are presented in Section~\ref{sec:simulations}. Conclusions and future research directions are discussed in Section~\ref{sec:conclusion}.

%%%%%%%%%%%%%%%%%%%%%%%%%%%%%%%%%%%%%%%%%%%%%%%%%%%%%%%%%%%%%%%%%%%%%%%%%%%%%%%%
\section{CONTROL ALGORITHM OVERVIEW}
\label{sec:overview}
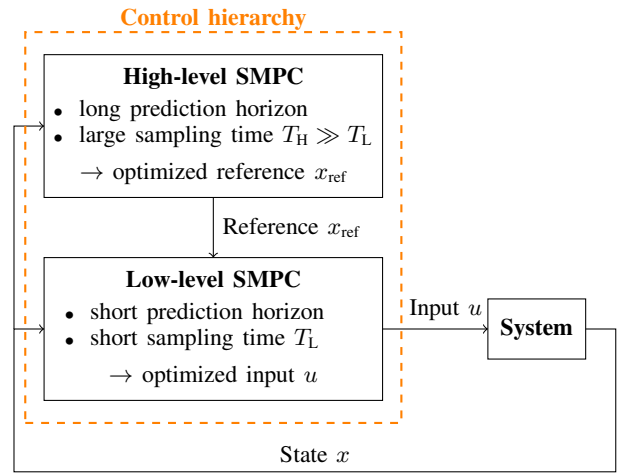
\begin{figure}
\vspace{0.1cm}
    \centering
    \begin{tikzpicture}

\def \sysx {4.3}
\def \sysy {0}
\def \cLLx {0}
\def \cLLy {0}
\def \cHLx {0}
\def \cHLy {2.7}
\def \lsys {1.3}
\def \hsys {0.8}
\def \lbx {4.5}
\def \hbx {1.9}
\def \arstp {0.4} % arrow step
\def \brstp {0.5} % used to shape larger rectangle length
\def \crstp {0.7} % used to shape larger rectangle height

\draw (\sysx-\lsys/2,\sysy-\hsys/2) rectangle ++(\lsys,\hsys) node[pos=0.5]{\small \textbf{System}};

\draw (\cLLx-\lbx/2,\cLLy-\hbx/2) rectangle ++(\lbx,\hbx) node[pos=0.5,text width=4.1cm,align=center]{\small \textbf{Low-level SMPC}\begin{itemize}
{\setlength\itemindent{-8pt} \item short prediction horizon}
{\setlength\itemindent{-8pt} \item short sampling time $T_\text{L}$}
\end{itemize}
$\rightarrow$ optimized input $u$};

\draw (\cHLx-\lbx/2,\cHLy-\hbx/2) rectangle ++(\lbx,\hbx) node[pos=0.5,text width=4.4cm,align=center]{\small \textbf{High-level SMPC}\begin{itemize}
{\setlength\itemindent{-8pt} \item long prediction horizon}
{\setlength\itemindent{-8pt} \item large sampling time $T_\text{H}\gg T_\text{L}$}
\end{itemize}
$\rightarrow$ optimized reference $x_\text{ref}$};

\draw[->] (\cLLx+\lbx/2,\cLLy) to node[pos=0.6,above] {\small Input $u$} (\sysx-\lsys/2,\sysy);

\draw[->] (\sysx+\lsys/2,\sysy) to (\sysx+\lsys/2+\arstp,\sysy) to (\sysx+\lsys/2+\arstp,\sysy-\hbx) to node[pos=0.5,above] {\small State $x$} (\cLLx-\lbx/2-\arstp,\sysy-\hbx) to (\cLLx-\lbx/2-\arstp,\cLLy) to (\cLLx-\lbx/2,\cLLy);

\draw[->] (\cLLx-\lbx/2-\arstp,\sysy) to (\cHLx-\lbx/2-\arstp,\cHLy) to (\cHLx-\lbx/2,\cHLy);

\draw[->] (\cHLx,\cHLy-\hbx/2) to node[pos=0.5,right] {\small Reference $x_\text{ref}$} (\cLLx,\cLLy+\hbx/2);

\draw [thick,dashed,orange](\cLLx/2+\cHLx/2-\lbx/2-\brstp/2,\cLLy/2+\cHLy/2-\hbx-\crstp) rectangle ++(\lbx+\brstp,2*\hbx+2*\crstp) {};

\node[above,orange] at (\cHLx,\cHLy+\hbx/2+\arstp/2){\small \textbf{Control hierarchy}};

\end{tikzpicture}
    \caption{Scheme of the control hierarchy.}
    \label{fig:controlhierarchyoverview}
\end{figure}

In this section, an overview of the control hierarchy is given, highlighting the role of each layer, and the connections between them. The proposed control algorithm is composed of two layers. Both the layers work in a receding horizon setting, selecting the optimal sequence of control actions by minimizing a properly designed cost function, while meeting the requirements formulated as a set of constraints. A scheme of the control hierarchy is represented in Fig.~\ref{fig:controlhierarchyoverview}.

The lower layer determines the immediate control actions to be applied on the system. A detailed model is used to compute the new optimal control input every time a new measurement of the state is available, and the constraints guarantee safety up to the user-defined probability.

The high-level layer is an additional structure, used to benefit from a long-term view in the optimization. The goal of this layer is not to determine the control input to be directly applied on the system, but rather to benefit from a long prediction horizon and provide this information to the low-level controller. For this reason, the optimization variable of this level is the reference trajectory used by the low-level controller, which is determined considering a longer prediction horizon. Moreover, since the reference trajectory must not be updated too often, a longer sampling time is used. Furthermore, the model used by the high-level controller is chosen to be an abstraction of the low-level one, including only the information that can be predicted in a  sufficiently reliable way even in the long term. Also, not necessarily all the state features (e.g. the lateral displacement) are relevant to decide the maneuver. Finally, the cost function and the constraints are based on the low-level ones, but adapted to be consistent with the high-level model.

%%%%%%%%%%%%%%%%%%%%%%%%%%%%%%%%%%%%%%%%%%%%%%%%%%%%%%%%%%%%%%%%%%%%%%%%%%%%%%%%
\section{DYNAMICAL MODELS}
\label{sec:dynamicalmodels}
This section introduces the dynamical model used by the SMPC algorithm to plan the ego vehicle (EV) trajectory, together with the dynamical models of the target vehicles (TVs) and of the pedestrians, which are used to predict future states and to generate safety constraints accordingly.

%%%%%%%%%%%%%%%%%%%%%%%%%%%%%%%%%%%%%%%%%%%%%%%%%%%%%%%%%%%%%%%%%%%%%%%%%%%%%%%%
\subsection{Ego Vehicle Model}
To predict the EV states within the prediction horizon, we used a kinematic bicycle model, expressed in the road-aligned reference frame of a (possibly curved) reference path, assumed known. The EV state is defined as $\bm{\xi} = [s,d,\phi,v]^\top$, where $(s,d)$ are the longitudinal and lateral position of the center of mass of the vehicle along the road, respectively, $\phi$ is the orientation of the vehicle with respect to the road itself, and $v$ is the linear velocity of the EV. The input vector is $\bm{u} = [a,\delta]^\top$, with acceleration $a$ and front steering angle $\delta$. Let $l_\text{f}$ and $l_\text{r}$ be the distances from the front and rear axles, respectively, to the center of gravity. Moreover, let $\kappa(s)$ be the local curvature of the reference path expressed as a function of the curvilinear coordinate $s$.

The nonlinear continuous-time dynamics is~\cite{CurvedBicycleDynamics}
\begin{equation}
    \dot{\bm{\xi}} =
    \begin{bmatrix}
    \displaystyle\frac{v\cos(\alpha+\phi)}{1-\kappa(s)d}\\[2ex]
    v\sin(\alpha+\phi)\\[2ex]
    \displaystyle v\bigg(\frac{\sin\alpha}{l_\text{r}}-\frac{\kappa(s)\cos(\alpha+\phi)}{1-\kappa(s)d}\bigg)\\[2ex]
    a
    \end{bmatrix}
    =:\bm{f}(\bm{\xi},\bm{u}),
    \label{eqn:continuousbicycle}
\end{equation}
where $\alpha=\arctan\bigg(\frac{l_\text{r}}{l_\text{f}+l_\text{r}}\tan\delta\bigg)$.
\begin{assumption}
During the motion, the spatial state components $(s,d)$ are such that $d\neq \frac{1}{\kappa(s)}$.
\label{ass:dassumption}
\end{assumption}

Assumption~\ref{ass:dassumption} is needed to avoid singularity issues, and it is realistic, as the curvature $\kappa$ of a road is typically small, and the controller keeps the lateral displacement $d$ limited.

A discrete-time linear model is obtained from~\eqref{eqn:continuousbicycle} along the lines of~\cite{BruedigamEtAl2021}. At first, the nonlinear dynamics is linearized around the current state $\bm{\xi}_0 = [s_0,d_0,\phi_0,v_0]^\top$ and the zero input $\bm{u}_0  = [0,0]^\top$, resulting in
\begin{equation}
    \dot{\bm{\xi}} = \bm{f}(\bm{\xi}_0,\bm{u}_0)+\bm{A}_\text{l}(\bm{\xi}-\bm{\xi}_0)+\bm{B}_\text{l}\bm{u},
\end{equation}
where
\begin{equation}
    \bm{A}_\text{l}=\frac{\partial\bm{f}}{\partial\bm{\xi}}\bigg\rvert_{(\bm{\xi}_0,\bm{u}_0)},\quad \bm{B}_\text{l}=\frac{\partial\bm{f}}{\partial\bm{u}}\bigg\rvert_{(\bm{\xi}_0,\bm{u}_0)}.
\end{equation}
Then, the discrete-time model is obtained as:
\begin{equation}
\begin{split}
    \bm{\xi}_{k+1}=&\ \bm{f}^\text{d}(\bm{\xi}_0,\bm{\xi}_k,\bm{u}_k)\\
    =&\ \bm{\xi}_0+\bm{f}(\bm{\xi}_0,\bm{u}_0)T+\bm{A}_\text{d}(\bm{\xi}_{k}-\bm{\xi}_0)+\bm{B}_\text{d}\bm{u}_{k},
    \label{eqn:discretelinearEV}
\end{split}
\end{equation}
where $(\bm{A}_\text{d},\bm{B}_\text{d})$ is obtained from $(\bm{A}_\text{l},\bm{B}_\text{l})$ with zero-order hold of sampling time $T$. In order to simplify the computation, the discretization is performed under the approximation that $k^\prime(s_0)=0$, where $k^\prime$ is the derivative of the path curvature with respect to the longitudinal position $s$.

Moreover, the EV is subject to the following constraints
\begin{subequations}
\begin{align}
    \left|d_k\right|\leq&\frac{w_\text{lane}}{2}-\frac{w_\text{veh}}{2}\label{eqn:lanebounds}\\
    0\leq&v_k\leq v_\text{max}\label{eqn:maxspeedLL}\\
    \bm{u}_\text{min}\leq&\bm{u}_k\leq\bm{u}_\text{max}\\
    \Delta\bm{u}_\text{min}\leq&\Delta\bm{u}_k\leq\Delta\bm{u}_\text{max},
\end{align}
\label{eqn:stateinputEVconstraints}%
\end{subequations}
where $\Delta\bm{u}_k=\bm{u}_k-\bm{u}_{k-1}$, and $w_\text{lane}$ and $w_\text{veh}$ are the lane and vehicle width, respectively.~\eqref{eqn:lanebounds} is formulated to ensure the EV shape never exceeds the lane bounds. Constraints~\eqref{eqn:stateinputEVconstraints} yield the set of admissible states $\Xi$ and inputs $\mathcal{U}$.
%%%%%%%%%%%%%%%%%%%%%%%%%%%%%%%%%%%%%%%%%%%%%%%%%%%%%%%%%%%%%%%%%%%%%%%%%%%%%%%%
\subsection{Target Vehicle Model}
In order to generate the safety constraints to avoid collisions, a prediction of the future positions $\bm{\xi}^\text{TV}_k$ of the TVs is needed. We use a discrete-time point-mass model~\cite{MizushimaEtAl2019}
\begin{align}
    \bm{\xi}^\text{TV}_{k+1} =&\ \bm{A}\bm{\xi}^\text{TV}_k+\bm{B}\bm{u}^\text{TV}_k\label{eqn:TVmodel}\\
    \bm{u}^\text{TV}_k =&\ \tilde{\bm{u}}^\text{TV}_k+\bm{w}^\text{TV}_k,
\end{align}
where $\bm{\xi}^\text{TV}=[x^\text{TV},v_x^\text{TV},y^\text{TV},v_y^\text{TV}]^\top$ is the TV state, composed of longitudinal position $x^\text{TV}$ and velocity $v_x^\text{TV}$, and lateral position $y^\text{TV}$ and velocity $v_y^\text{TV}$, expressed in the inertial world frame, and $\bm{u}^\text{TV}$ is the system input, consisting of the longitudinal and lateral accelerations. The TV matrices are
\begin{equation}
    \bm{A} =
    \begin{bmatrix}
    1 & T & 0 & 0\\
    0 & 1 & 0 & 0\\
    0 & 0 & 1 & T\\
    0 & 0 & 0 & 1
    \end{bmatrix}
    ,\quad \bm{B} =
    \begin{bmatrix}
    0.5T^2 & 0\\
    T & 0\\
    0 & 0.5T^2\\
    0 & T
    \end{bmatrix}
    .
    \label{eqn:ABTVP}
\end{equation}

We assume the input to be the sum of a feedback term $\tilde{\bm{u}}^\text{TV}$ and a perturbation term $\bm{w}^\text{TV}_k$. The feedback term is designed to correct possible deviations from the desired trajectory $\bm{\xi}^\text{TV}_\text{ref}$, and is given by
\begin{equation}
   \tilde{\bm{u}}^\text{TV}_k= \bm{K}\Delta\bm{\xi}_k^\text{TV}
   \label{eqn:TVfeedbackgainLL},
\end{equation}
where $\Delta\bm{\xi}_k^\text{TV} = (\bm{\xi}^\text{TV}_k-\bm{\xi}^\text{TV}_{\text{ref},k})$, and $\bm{K}$ is designed according to a linear quadratic control strategy, where we assume the longitudinal reference trajectory is only given in terms of velocity. To model the uncertainty of the future TV motion, we assume $\bm{w}^\text{TV}_k$ to be independent zero-mean Gaussian noises with covariance matrix $\bm{\Sigma}^\text{TV}_{\bm{w}}$. Moreover, we assume that the TV input $\bm{u}^\text{TV}$ is saturated between $\bm{u}^\text{TV}_\text{min}$ and $\bm{u}^\text{TV}_\text{max}$.
%%%%%%%%%%%%%%%%%%%%%%%%%%%%%%%%%%%%%%%%%%%%%%%%%%%%%%%%%%%%%%%%%%%%%%%%%%%%%%%%
\subsection{Pedestrian Model}
To compute the prediction of the future states of the pedestrians, we refer to the model in~\cite{MuraleedharanEtAl2020}, which reflects the model used for the TV. The pedestrian state is $\bm{\xi}^\text{P}=[x^\text{P},v_x^\text{P},y^\text{P},v_y^\text{P}]^\top$, and the dynamics is
\begin{equation}
    \bm{\xi}^\text{P}_{k+1} = \bm{A}\bm{\xi}^\text{P}_k+\bm{B}\bm{w}^\text{P}_k,
    \label{eqn:Pmodel}
\end{equation}
where $\bm{A}$ and $\bm{B}$ are as in~\eqref{eqn:ABTVP}. The pedestrian noise $\bm{w}^\text{P}_k$ is an independent zero-mean Gaussian noise with covariance matrix $\bm{\Sigma}^\text{P}_{\bm{w}}$. Observe that the pedestrian input is modeled as pure noise, with no stabilizing feedback loop.

%%%%%%%%%%%%%%%%%%%%%%%%%%%%%%%%%%%%%%%%%%%%%%%%%%%%%%%%%%%%%%%%%%%%%%%%%%%%%%%%
\section{TRAJECTORY PLANNING}
\label{sec:trajectoryplanning}
In the following, the details of the low-level SMPC algorithm used for the trajectory planning are discussed.

%%%%%%%%%%%%%%%%%%%%%%%%%%%%%%%%%%%%%%%%%%%%%%%%%%%%%%%%%%%%%%%%%%%%%%%%%%%%%%%%
\subsection{Safety Distance}
\label{subsec:safetydistanceLL}
In this section, a safety distance to be observed from each traffic participant is derived, so that the collision is avoided at least with the required probability $\beta$. Considering the TV, along the lines of~\cite{BruedigamEtAl2021}, the safety distance can be obtained as

\begin{equation}
    a_k^\text{TV} = \frac{l_\text{veh}^\text{TV}}{2}+\Delta s_\text{stop}^\text{TV}+e_{s,k}^\text{TV}+\epsilon_\text{safe}^\text{TV},
\end{equation}
with $l_\text{veh}^\text{TV}$ the TV length, and $\Delta s_\text{stop}^\text{TV}$ the distance needed to stop if the EV is driving faster.  $e_{s,k}^\text{TV}$ takes into account the uncertainty on the TV position, and $\epsilon_\text{safe}^\text{TV}$ is an extra safety parameter. $a_k^\text{TV}$ is computed for each prediction step $k$. In the following, the superscript TV is omitted for readability.

The term $e_{s,k}$ deterministically increases the safety distance such that the considered region contains the TV position with probability $\beta$. Given the linear model~\eqref{eqn:TVmodel}, and the Gaussian distribution of the noise $\bm{w}_k$, the prediction error covariance matrix $\bm{\Sigma}^e_k$ is obtained propagating the uncertainty
\begin{equation}
    \bm{\Sigma}^e_{k+1} = \bm{B}\bm{\Sigma}_{\bm{w}}\bm{B}^\top+(\bm{A}+\bm{B}\bm{K})\bm{\Sigma}^e_k(\bm{A}+\bm{B}\bm{K})^\top,
\end{equation}
starting from the initialization $\bm{\Sigma}^e_0=\bm{0}$ (the current TV position is assumed to be perfectly known). From $\bm{\Sigma}^e_k$, we obtain the standard deviation of the longitudinal position prediction error $\sigma_{s,k}$, and, using the standard approximation explained in detail in~\cite{BruedigamEtAl2021}, the term $e_{s,k}=\sigma_{s,k}\sqrt\gamma$ is derived, where $\gamma=-2\ln(1-\beta)$. The same procedure is repeated to obtain the safety distance from the pedestrian $a_k^\text{P}$.

Note that, even though $a_k^\text{TV}$ and $a_k^\text{P}$ have been derived in a probabilistic framework, they determine a deterministic \textit{safety area} around the predicted position of each other traffic participant. As a result, the chance constraints can be reformulated as deterministic constraints preventing the EV from entering the safety areas, as explained in Section~\ref{subsec:LLconstraints}.

%%%%%%%%%%%%%%%%%%%%%%%%%%%%%%%%%%%%%%%%%%%%%%%%%%%%%%%%%%%%%%%%%%%%%%%%%%%%%%%%
\subsection{Safety Constraints}
\label{subsec:LLconstraints}
\begin{figure}
\vspace{0.1cm}
    \centering
    \scalebox{1.15}{
    \begin{tikzpicture}

\def \wlane {0.25}
\def \leftlim {-1.1}
\def \rightlim {1.1}
\def \uplim {0.5}
\def \lowmin {-0.5}
\def \upmax {0.6}
\def \txtofst {0.15}
\def \scsp {2.5}  %   scene space
\def \lveh {0.4}
\def \wveh {0.15}
\def \sped {0.15} % pedestrian side
\def \evx {-0.75}
\def \tvx {0.4}
\def \pdx {0.4}
\def \pdy {-0.1}
\def \dist {0.2}
\def \arlen {0.2}

%   TV same lane
\draw[draw=red,thick] (\tvx-\lveh/2-\dist-\scsp,-2*\wlane) to (\tvx-\lveh/2-\dist-\scsp,\uplim);
\path [pattern=custom north east lines,hatchspread=5pt,hatchcolor=red!50] (\tvx-\lveh/2-\dist-\scsp,-2*\wlane) -- (\tvx-\lveh/2-\dist-\scsp,\uplim) -- (\rightlim-\scsp,\uplim) -- (\rightlim-\scsp,-2*\wlane);

\draw (-\scsp/2,\lowmin) to (-\scsp/2,\upmax);
\node at (-\scsp,\upmax+\txtofst) {\small 1) TV same lane};

\draw[dashed](\leftlim-\scsp,0) to (\rightlim-\scsp,0);
\draw[thick](\leftlim-\scsp,-\wlane) to (\rightlim-\scsp,-\wlane);
\draw[thick](\leftlim-\scsp,\wlane) to (\rightlim-\scsp,\wlane);

\draw [thick,draw=red,fill=red!20] (\evx-\lveh/2-\scsp,-\wlane/2-\wveh/2) rectangle ++(\lveh,\wveh) node[pos=.5]{};
\draw[->,thick,draw=red](\evx+\lveh/2-\scsp,-\wlane/2) to (\evx+\lveh/2-\scsp+\arlen,-\wlane/2);

\draw [thick,draw=blue,fill=blue!20] (\tvx-\lveh/2-\scsp,-\wlane/2-\wveh/2) rectangle ++(\lveh,\wveh)node[pos=.5]{};
\draw[->,thick,draw=blue](\tvx+\lveh/2-\scsp,-\wlane/2) to (\tvx+\lveh/2-\scsp+\arlen,-\wlane/2);

%   TV intersection
\draw[draw=red,thick] (-1.3*\wlane,-2*\wlane) to (-1.3*\wlane,\uplim);
\path [pattern=custom north east lines,hatchspread=5pt,hatchcolor=red!50] (-1.3*\wlane,-2*\wlane) -- (-1.3*\wlane,\uplim) -- (\rightlim,\uplim) -- (\rightlim,-2*\wlane);

\node at (0,\upmax+\txtofst) {\small 2) TV intersection};

\draw[dashed](\leftlim,0) to (\rightlim,0);
\draw[thick](\leftlim,-\wlane) to (\rightlim,-\wlane);
\draw[thick](\leftlim,\wlane) to (-\wlane,\wlane);
\draw[thick](\wlane,\wlane) to (\rightlim,\wlane);
\draw[thick](-\wlane,\wlane) to (-\wlane,\uplim);
\draw[thick](\wlane,\wlane) to (\wlane,\uplim);
\draw[dashed](0,\wlane) to (0,\uplim);

\draw [thick,draw=red,fill=red!20] (\evx-\lveh/2,-\wlane/2-\wveh/2) rectangle ++(\lveh,\wveh) node[pos=.5]{};
\draw[->,thick,draw=red](\evx+\lveh/2,-\wlane/2) to (\evx+\lveh/2+\arlen,-\wlane/2);

\draw [thick,draw=blue,fill=blue!20] (\tvx-\lveh/2,\wlane/2-\wveh/2) rectangle ++(\lveh,\wveh)node[pos=.5]{};
\draw[->,thick,draw=blue](\tvx-\lveh/2,\wlane/2) to (\tvx-\lveh/2-\arlen,\wlane/2);

%   Pedestrian
\draw[draw=red,thick] (\pdx-\sped/2-\dist+\scsp,-2*\wlane) to (\pdx-\sped/2-\dist+\scsp,\uplim);
\path [thin,pattern=custom north east lines,hatchspread=5pt,hatchcolor=red!50] (\pdx-\sped/2-\dist+\scsp,-2*\wlane) -- (\pdx-\sped/2-\dist+\scsp,\uplim) -- (\rightlim+\scsp,\uplim) -- (\rightlim+\scsp,-2*\wlane);
%\path [thin,pattern=north east lines,pattern color=red!50] (\pdx-\sped/2-\dist+\scsp,-2*\wlane) -- (\pdx-\sped/2-\dist+\scsp,\uplim) -- (\rightlim+\scsp,\uplim) -- (\rightlim+\scsp,-2*\wlane);

\draw (\scsp/2,\lowmin) to (\scsp/2,\upmax);
\node at (\scsp,\upmax+\txtofst) {\small 3) Pedestrian};

\draw[dashed](\leftlim+\scsp,0) to (\rightlim+\scsp,0);
\draw[thick](\leftlim+\scsp,-\wlane) to (\rightlim+\scsp,-\wlane);
\draw[thick](\leftlim+\scsp,\wlane) to (\rightlim+\scsp,\wlane);

\draw [thick,draw=red,fill=red!20] (\evx-\lveh/2+\scsp,-\wlane/2-\wveh/2) rectangle ++(\lveh,\wveh) node[pos=.5]{};
\draw[->,thick,draw=red](\evx+\lveh/2+\scsp,-\wlane/2) to (\evx+\lveh/2+\scsp+\arlen,-\wlane/2);

\draw [thick,draw=pedestriangreen,fill=pedestriangreen!20] (\pdx-\sped/2+\scsp,\pdy-\sped/2) rectangle ++(\sped,\sped)node[pos=.5]{};
\draw[->,thick,draw=pedestriangreen](\pdx+\scsp,\pdy+\sped/2) to (\pdx+\scsp,\pdy+\sped/2+\arlen);

\end{tikzpicture}
    }
    \caption{Constraints used in the low-level trajectory planner.}
    \label{fig:LLconstraints}
\end{figure}
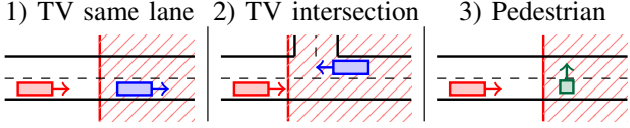

In this section, we derive the deterministic constraints to ensure the EV does not enter the safety areas of the other traffic participants derived in section~\ref{subsec:safetydistanceLL}. Each possible scenario is considered separately, and the constraints are represented in Fig.~\ref{fig:LLconstraints}. The constraints are derived considering a simple urban environment composed of roads with a single lane (per direction) and an intersection. Moreover, it is assumed the EV can measure the state of other traffic participants in road-aligned coordinates. Note that a new constraint is generated for each traffic participant.
%%%%%%%%%%%%%%%%%%%%%%%%%%%%%%%%%%%%%%%%%%%%%%%%%%%%%%%%%%%%%%%%%%%%%%%%%%%%%%%%
\subsubsection{TV in same lane}
If a TV is in the same lane in front, the EV must stay behind. 
Thus, the following constraint is generated, ensuring the safety distance $a_k^\text{TV}$ is always kept
\begin{equation}
    s_k+\frac{l_\text{veh}}{2}\leq s_k^\text{TV}-a_k^\text{TV}.
    \label{eqn:LLTVsamelane}
\end{equation}
%%%%%%%%%%%%%%%%%%%%%%%%%%%%%%%%%%%%%%%%%%%%%%%%%%%%%%%%%%%%%%%%%%%%%%%%%%%%%%%%
\subsubsection{TV at intersection}
If the EV is approaching the intersection, planning a left turn, and a TV is predicted to cross the intersection before the EV can leave the intersection free, then the EV must wait before entering the intersection, giving the right of way. As a result, a constraint is generated before the intersection area
\begin{equation}
    s_k\leq s_\text{int}.
    \label{eqn:LLTVintersection}
\end{equation}
%%%%%%%%%%%%%%%%%%%%%%%%%%%%%%%%%%%%%%%%%%%%%%%%%%%%%%%%%%%%%%%%%%%%%%%%%%%%%%%%
\subsubsection{Pedestrian}
When a pedestrian is predicted to cross the road, the EV must safely stop before. Thus, we have
\begin{equation}
    s_k+\frac{l_\text{veh}}{2}\leq s_k^\text{P}-a_k^\text{P}.
    \label{eqn:LLPedestrian}
\end{equation}
%%%%%%%%%%%%%%%%%%%%%%%%%%%%%%%%%%%%%%%%%%%%%%%%%%%%%%%%%%%%%%%%%%%%%%%%%%%%%%%%
\subsection{Optimal Control Problem}
Given the EV model~\eqref{eqn:discretelinearEV}, the state and input constraints from~\eqref{eqn:stateinputEVconstraints}, and the positional constraints derived in Section~\ref{subsec:LLconstraints}, the full optimal control problem for the low-level trajectory planner is formulated:
\begin{subequations}
\begin{equation}
    \min_{\bm{u}_k}\sum_{k=0}^{N-1}\|\Delta\bm{\xi}_k\|^2_{\bm{Q}}+\|\bm{u}_k\|^2_{\bm{R}}+\|\Delta\bm{u}_k\|^2_{\bm{S}}+\|\Delta\bm{\xi}_{N}\|^2_{\bm{P}}
    \label{eqn:LLcostfunction}
\end{equation}
\begin{alignat}{2}
    \text{s.t. }\bm{\xi}_{k+1}=&\ \bm{f}^\text{d}(\bm{\xi}_0,\bm{\xi}_k,\bm{u}_k)\ &&\forall k\in\{0,\dots,N-1\}\label{eqn:EVLLcompactmodel}\\
    \bm{\xi}^\text{TV}_{k+1} =&\ \bm{A}\bm{\xi}^\text{TV}_k+\bm{B}\bm{u}^\text{TV}_k\ &&\forall k\in\{0,\dots,N-1\}\\
    \bm{\xi}^\text{P}_{k+1} =&\ \bm{A}\bm{\xi}^\text{P}_k+\bm{B}\bm{w}^\text{P}_k\ &&\forall k\in\{0,\dots,N-1\}\\
    \bm{\xi}_k\in&\ \bm{\Xi}\ &&\forall k\in\{1,\dots,N\}\\
    \bm{u}_k\in&\ \mathcal{U}\ &&\forall k\in\{0,\dots,N-1\}\\
    0\geq&\ \bm{q}_s(\bm{\xi}_0,\bm{\xi}_k)s_k+\bm{q}_d&&(\bm{\xi}_0,\bm{\xi}_k)d_k+\bm{q}_t(\bm{\xi}_0,\bm{\xi}_k)\nonumber\\
    & &&\forall k\in\{1,\dots,N\},
    \label{eqn:positionalconstraintsOCPLL}
\end{alignat}
\label{prb:OCPLL}%
\end{subequations}
where $N$ is the prediction horizon, $\bm{Q}$, $\bm{R}$, $\bm{S}$, and $\bm{P}$ are positive semi-definite weighting matrices, and, for the input difference, $\bm{u}_{-1}$ is the last applied input.

Note that the positional constraint coefficients $\bm{q}_s$, $\bm{q}_d$, and $\bm{q}_t$ in~\eqref{eqn:positionalconstraintsOCPLL}, which are a reformulation of the constraints~\eqref{eqn:LLTVsamelane},~\eqref{eqn:LLTVintersection}, and~\eqref{eqn:LLPedestrian}, can be computed from the current EV state and the predicted TV and pedestrian states, before the optimization starts. As a result, the optimal control problem~\eqref{prb:OCPLL} is a quadratic program with linear constraints.
%%%%%%%%%%%%%%%%%%%%%%%%%%%%%%%%%%%%%%%%%%%%%%%%%%%%%%%%%%%%%%%%%%%%%%%%%%%%%%%%
\section{MANEUVER PLANNING}
\label{sec:maneuverplanning}
In this section, the high-level maneuver planner is introduced, specifying the motivation, the details of the model being used, the constraints generation, and the cost function.

%%%%%%%%%%%%%%%%%%%%%%%%%%%%%%%%%%%%%%%%%%%%%%%%%%%%%%%%%%%%%%%%%%%%%%%%%%%%%%%%
\subsection{High-Level Models}
\label{subsec:HLmodels}
The scope of the maneuver planner is to determine the optimal sequence of high-level actions, considering the future behavior of the other vehicles in a long prediction horizon. However, planning maneuvers in the long run requires fewer details with respect to the immediate trajectory planning for the short term. Moreover, due to the uncertainty of the behavior of the other traffic participants, a detailed prediction of other vehicles or pedestrians for farther steps would not be reliable. For this reason, here we present simplified models, which are used at the higher level to predict the only information which can be predicted reliably enough even in the long term, and which is needed to plan the maneuver.

Given the simple structure of the road, the main maneuver the EV can plan is to accelerate or decelerate, for example in order to reach the intersection when it is free, avoiding a full stop that might otherwise be required. In order to determine the optimal speed, we only consider the position of the EV along the (fixed) reference path, which evolves according to a point-mass model dynamics with piece-wise constant speed, without paying attention to lateral displacements. As a result, the high-level dynamics is
\begin{equation}
    s_{h+1} = s_h+\nu_h T_\text{H},
    \label{eqn:HLmodel}
\end{equation}
where the input $\nu$ is the speed, and $h$ is the prediction step of the high-level maneuver planner. $T_\text{H}$ and $N_\text{H}$ are the sampling time and the prediction horizon of the high-level model, respectively, and we assume $N_\text{H}T_\text{H}\geq NT$. The simplified model is subject to input constraints in the form
\begin{equation}
    0\leq\nu_h\leq\nu_\text{max}(s_h).
    \label{eqn:HLmaxspeed}
\end{equation}
$\nu_\text{max}(s_h)$ is the maximum speed allowed depending on the position $s_h$, and might be lower than $v_\text{max}$ in~\eqref{eqn:maxspeedLL}, e.g., to take into account that it is unrealistic to drive at the maximum speed in highly curved stretches. This constraint makes the prediction more reliable, since the simplified model does not take the path's curvature into account.

Concerning the other traffic participants, it is enough to predict their longitudinal position along the path, in order to formulate the constraints of Section~\ref{subsec:HLconstraints}. This can be obtained by computing the equivalent of models~\eqref{eqn:TVmodel} and~\eqref{eqn:Pmodel} with a time-step $T_\text{H}$, and by projecting the position of the TV and the pedestrian on the reference path, obtaining $s_h^\text{TV}$ and $s_h^\text{P}$. Firstly, the point-mass linear model matrices~\eqref{eqn:ABTVP} are recomputed using $T_\text{H}$ in place of $T$, obtaining $\bm A_\text{H}$ and $\bm B_\text{H}$. The feedback gain $\bm K$ from~\eqref{eqn:TVfeedbackgainLL} is also substituted by $\bm K_\text{H}$, which is computed so that the closed-loop matrix $\bm A_\text{H}+\bm B_\text{H}\bm K_\text{H}$ is the equivalent of $\bm A+\bm B\bm K$ with a sampling time $T_\text{H}$ instead of $T$. Then, an appropriate noise is obtained, considering the average of the $\bar{k}=\lfloor\frac{T_\text{H}}{T}\rfloor$ noises $\bm{w}^\text{TV}_k,\dots,\bm{w}^\text{TV}_{k+\bar{k}-1}$ affecting the TV within a high-level sampling time $T_\text{H}$. The average noise $\bm{w}^\text{TV}_h$ is still a zero-mean Gaussian, and, being $\bm{w}^\text{TV}_k$ independent, its variance is
\begin{equation}
    \text{Var}\bigg(\sum_{j=0}^{\bar{k}-1}\frac{\bm{w}^\text{TV}_{k+j}}{\bar{k}}\bigg)=\sum_{j=0}^{\bar{k}-1}\text{Var}\bigg(\frac{\bm{w}^\text{TV}_{k+j}}{\bar{k}}\bigg)=\frac{\bm{\Sigma}^\text{TV}_{\bm{w}}}{\bar{k}}.
\end{equation}
The same procedure is repeated for the pedestrian noise $\bm{w}^\text{P}_h$.

%%%%%%%%%%%%%%%%%%%%%%%%%%%%%%%%%%%%%%%%%%%%%%%%%%%%%%%%%%%%%%%%%%%%%%%%%%%%%%%%
\subsection{High-Level Constraints}
\label{subsec:HLconstraints}
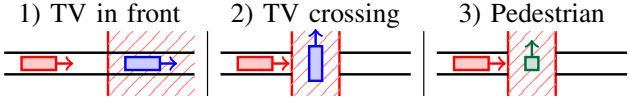
\begin{figure}
\vspace{0.1cm}
    \centering
    \scalebox{1.15}{
    \begin{tikzpicture}

\def \wlane {0.25}
\def \leftlim {-1.1}
\def \rightlim {1.1}
\def \uplim {0.8}
\def \lowmin {-0.5}
\def \upmax {0.25}
\def \txtofst {0.18}
\def \scsp {2.5}  %   scene space
\def \lveh {0.4}
\def \wveh {0.15}
\def \sped {0.15} % pedestrian side
\def \evx {-0.7}
\def \tvx {0.5}
\def \pdx {0}
\def \pdy {-0.125}
\def \dist {0.2}
\def \arlen {0.2}

%   TV in front
\draw[draw=red,thick] (\tvx-\lveh/2-\dist-\scsp,-2*\wlane) to (\tvx-\lveh/2-\dist-\scsp,\wlane);
\path [pattern=custom north east lines,hatchspread=5pt,hatchcolor=red!50] (\tvx-\lveh/2-\dist-\scsp,-2*\wlane) -- (\tvx-\lveh/2-\dist-\scsp,\wlane) -- (\rightlim-\scsp,\wlane) -- (\rightlim-\scsp,-2*\wlane);

\draw (-\scsp/2,\lowmin) to (-\scsp/2,\upmax);
\node at (-\scsp,\upmax+\txtofst) {\small 1) TV in front};

\draw[thick](\leftlim-\scsp,0) to (\rightlim-\scsp,0);
\draw[thick](\leftlim-\scsp,-\wlane) to (\rightlim-\scsp,-\wlane);

\draw [thick,draw=red,fill=red!20] (\evx-\lveh/2-\scsp,-\wlane/2-\wveh/2) rectangle ++(\lveh,\wveh) node[pos=.5]{};
\draw[->,thick,draw=red](\evx+\lveh/2-\scsp,-\wlane/2) to (\evx+\lveh/2-\scsp+\arlen,-\wlane/2);

\draw [thick,draw=blue,fill=blue!20] (\tvx-\lveh/2-\scsp,-\wlane/2-\wveh/2) rectangle ++(\lveh,\wveh)node[pos=.5]{};
\draw[->,thick,draw=blue](\tvx+\lveh/2-\scsp,-\wlane/2) to (\tvx+\lveh/2-\scsp+\arlen,-\wlane/2);

%   TV crossing
\draw[draw=red,thick] (-\wveh/2-\dist,-2*\wlane) to (-\wveh/2-\dist,\wlane);
\draw[draw=red,thick] (\wveh/2+\dist,-2*\wlane) to (\wveh/2+\dist,\wlane);
\path [pattern=custom north east lines,hatchspread=5pt,hatchcolor=red!50] (-\wveh/2-\dist,-2*\wlane) -- (-\wveh/2-\dist,\wlane) -- (\wveh/2+\dist,\wlane) -- (\wveh/2+\dist,-2*\wlane);

\node at (0,\upmax+\txtofst) {\small 2) TV crossing};

\draw[thick](\leftlim,0) to (-\dist-\wveh/2,0);
\draw[thick](\dist+\wveh/2,0) to (\rightlim,0);
\draw[thick](\leftlim,-\wlane) to (-\dist-\wveh/2,-\wlane);
\draw[thick](\dist+\wveh/2,-\wlane) to (\rightlim,-\wlane);

\draw [thick,draw=red,fill=red!20] (\evx-\lveh/2,-\wlane/2-\wveh/2) rectangle ++(\lveh,\wveh) node[pos=.5]{};
\draw[->,thick,draw=red](\evx+\lveh/2,-\wlane/2) to (\evx+\lveh/2+\arlen,-\wlane/2);

\draw [thick,draw=blue,fill=blue!20] (-\wveh/2,-\wlane/2-\lveh/2) rectangle ++(\wveh,\lveh)node[pos=.5]{};
\draw[->,thick,draw=blue](0,-\wlane/2+\lveh/2) to (0,-\wlane/2+\lveh/2+\arlen);

%   Pedestrian
\draw[draw=red,thick] (\pdx-\sped/2-\dist+\scsp,-2*\wlane) to (\pdx-\sped/2-\dist+\scsp,\wlane);
\draw[draw=red,thick] (\pdx+\sped/2+\dist+\scsp,-2*\wlane) to (\pdx+\sped/2+\dist+\scsp,\wlane);
\path [pattern=custom north east lines,hatchspread=5pt,hatchcolor=red!50] (\pdx-\sped/2-\dist+\scsp,-2*\wlane) -- (\pdx-\sped/2-\dist+\scsp,\wlane) -- (\pdx+\sped/2+\dist+\scsp,\wlane) -- (\pdx+\sped/2+\dist+\scsp,-2*\wlane);

\draw (\scsp/2,\lowmin) to (\scsp/2,\upmax);
\node at (\scsp,\upmax+\txtofst) {\small 3) Pedestrian};

\draw[thick](\leftlim+\scsp,0) to (-\dist-\sped/2+\scsp,0);
\draw[thick](\dist+\sped/2+\scsp,0) to (\rightlim+\scsp,0);
\draw[thick](\leftlim+\scsp,-\wlane) to (-\dist-\sped/2+\scsp,-\wlane);
\draw[thick](\dist+\sped/2+\scsp,-\wlane) to (\rightlim+\scsp,-\wlane);

\draw [thick,draw=red,fill=red!20] (\evx-\lveh/2+\scsp,-\wlane/2-\wveh/2) rectangle ++(\lveh,\wveh) node[pos=.5]{};
\draw[->,thick,draw=red](\evx+\lveh/2+\scsp,-\wlane/2) to (\evx+\lveh/2+\scsp+\arlen,-\wlane/2);

\draw [thick,draw=pedestriangreen,fill=pedestriangreen!20] (\pdx-\sped/2+\scsp,\pdy-\sped/2) rectangle ++(\sped,\sped)node[pos=.5]{};
\draw[->,thick,draw=pedestriangreen](\pdx+\scsp,\pdy+\sped/2) to (\pdx+\scsp,\pdy+\sped/2+\arlen);

\end{tikzpicture}
    }
    \caption{Constraints used in the high-level maneuver planner. Coherently with the abstract model~\eqref{eqn:HLmodel}, no lateral displacement (or lane) is considered.}
    \label{fig:HLconstraints}
\end{figure}

In this section, the constraints of the high-level optimal control problem are formulated. Repeating the procedure of Section~\ref{subsec:safetydistanceLL}, the safety distances $\rho_h^\text{TV}$ and $\rho_h^\text{P}$ are derived, based on the statistic description of the high-level noises $\bm{w}^\text{TV}_h$ and $\bm{w}^\text{P}_h$, respectively, and on the risk parameter $\beta_\text{H}$. Similarly to the low-level case in Section~\ref{subsec:LLconstraints}, each case is considered separately and a proper constraint is generated. A scheme of the constraints is given in Fig.~\ref{fig:HLconstraints}.

%%%%%%%%%%%%%%%%%%%%%%%%%%%%%%%%%%%%%%
\subsubsection{TV in front}
If a TV is predicted to occupy the same lane in front, a constraint is generated behind the TV
\begin{equation}
    s_h+\frac{l_\text{veh}}{2}\leq s_h^\text{TV}-\rho_h^\text{TV}.
    \label{eqn:HLTVfrontconstraint}
\end{equation}
Note that~\eqref{eqn:HLTVfrontconstraint} is not needed for safety, guaranteed by the analogous condition~\eqref{eqn:LLTVsamelane} in the low level, but rather to make the maneuver planner aware of the future position of the TV, allowing to plan a smooth deceleration when needed.
%%%%%%%%%%%%%%%%%%%%%%%%%%%%%%%%%%%%%%
\subsubsection{TV crossing the path}
When a TV is predicted to temporarily cross the path of the EV (e.g., at an intersection), the maneuver planner should take this into consideration and plan the speed accordingly. If possible, the EV could reach the intersection point before, when it is still free, or otherwise it could plan to decelerate, in order to smooth the motion and minimize the stop time.
The constraint is designed as
\begin{equation}
    \begin{cases}
    \displaystyle s_h+\frac{l_\text{veh}}{2}\leq s_h^\text{TV}-\rho_h^\text{TV}& \text{if } s_h\leq s_h^\text{TV}\\[2ex]
    \displaystyle s_h-\frac{l_\text{veh}}{2}\geq s_h^\text{TV}+\rho_h^\text{TV}& \text{if } s_h>s_h^\text{TV},
    \end{cases}
    \label{eqn:tempHLconstraintTVcrossing}
\end{equation}
which can be reformulated in a quadratic expression
\begin{equation}
    (s_h-s_h^\text{TV}+\Delta_1)(-s_h+s_h^\text{TV}+\Delta_2)\leq0,
    \label{eqn:HLconstraint1}
\end{equation}
where $\Delta_1$ and $\Delta_2$ collect the terms from~\eqref{eqn:tempHLconstraintTVcrossing}, plus possible additional safety parameters. Note that not only constraint~\eqref{eqn:HLconstraint1} is nonlinear, but it is also position dependent, and generates a non-connected admissible set.
%%%%%%%%%%%%%%%%%%%%%%%%%%%%%%%%%%%%%%
\subsubsection{Pedestrian}
Similar to the previous case, if a pedestrian is predicted to cross the street, the EV should plan to get over the crossing point before the pedestrian starts to cross, or to decelerate to minimize the stop time potentially required.
%%%%%%%%%%%%%%%%%%%%%%%%%%%%%%%%%%%%%%%%%%%%%%%%%%%%%%%%%%%%%%%%%%%%%%%%%%%%%%%%
\subsection{Optimal Control Problem}
From the simplified models outlined in Section~\ref{subsec:HLmodels}, and the constraints in Section~\ref{subsec:HLconstraints}, we can formulate the optimal control problem for the high-level maneuver planner:
\begin{subequations}
\begin{equation}
    \min_{\nu_h}\sum_{h=0}^{N_\text{H}-1}
    (\nu_h-\nu_{h-1})^2+r_H(\nu_h-v_\text{ref})^2
    \label{eqn:HLcostfunction}
\end{equation}
\begin{alignat}{2}
    \text{s.t. }s_{h+1} =&\ s_h+\nu_h T_\text{H}&&\forall h\in\{0,\dots,N_\text{H}-1\}\\
    \bm{\xi}^\text{TV}_{h+1} =&\ \bm{A}_\text{H}\bm{\xi}^\text{TV}_h+\bm{B}_\text{H}\bm{u}^\text{TV}_h&&\forall h\in\{0,\dots,N_\text{H}-1\}\\
    \bm{\xi}^\text{P}_{h+1} =&\ \bm{A}_\text{H}\bm{\xi}^\text{P}_h+\bm{B}_\text{H}\bm{w}^\text{P}_h&&\forall h\in\{0,\dots,N_\text{H}-1\}\\
    (s_{h+1},&\nu_h)\in\mathcal{Z}&&\forall h\in\{0,\dots,N_\text{H}-1\}\label{eqn:HLcompactconstraints}
\end{alignat}
\label{prb:OCPHL}%
\end{subequations}
where $\nu_{-1}$ is set equal to the current speed $v_0$, and set $\mathcal{Z}$ in~\eqref{eqn:HLcompactconstraints} compactly represents constraints~\eqref{eqn:HLmaxspeed},~\eqref{eqn:HLTVfrontconstraint}, and~\eqref{eqn:HLconstraint1}. The cost function~\eqref{eqn:HLcostfunction} is designed to penalize large accelerations, for sake of comfort and coherently with~\eqref{eqn:LLcostfunction}, and to penalize deviations from a desired cruise speed. This second term is needed to avoid that the optimal control problem~\eqref{prb:OCPHL} admits the trivial optimal solution $\nu_h=0\ \forall h$ when the current EV speed is zero. Note that the optimal reference speed $\nu_h$ must be updated solving~\eqref{prb:OCPHL} once every $T_\text{H}$ seconds, and not at each iteration of the low-level controller.

%%%%%%%%%%%%%%%%%%%%%%%%%%%%%%%%%%%%%%%%%%%%%%%%%%%%%%%%%%%%%%%%%%%%%%%%%%%%%%%%
\section{RESULTS}
\label{sec:simulations}
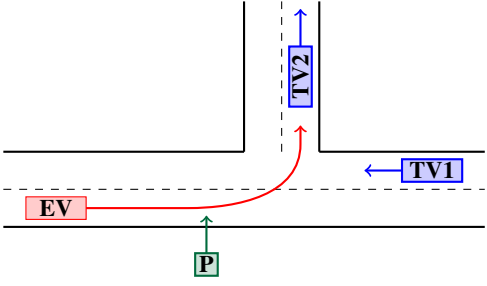
\begin{figure}
\vspace{0.1cm}
    \centering
    \begin{tikzpicture}

\def \wlane {0.5}
\def \leftlim {-3.7}
\def \rightlim {2.7}
\def \uplim {2.5}
\def \lveh {0.8}
\def \wveh {0.3}
\def \sped {0.3} %  pedestrian side
\def \arlen {0.5}
\def \evx {-3}
\def \tvx {2}
\def \tvy {1.5}
\def \pdx {-1}
\def \pdy {-1}

\draw[dashed](\leftlim,0) to (\rightlim,0);
\draw[thick](\leftlim,-\wlane) to (\rightlim,-\wlane);
\draw[thick](\leftlim,\wlane) to (-\wlane,\wlane);
\draw[thick](\wlane,\wlane) to (\rightlim,\wlane);
\draw[thick](-\wlane,\wlane) to (-\wlane,\uplim);
\draw[thick](\wlane,\wlane) to (\wlane,\uplim);
\draw[dashed](0,\wlane) to (0,\uplim);

\draw [draw=red,fill=red!20] (\evx-\lveh/2,-\wlane/2-\wveh/2) rectangle ++(\lveh,\wveh) node[pos=.5] {\small \textbf{EV}};
\draw[->,thick,draw=red](\evx+\lveh/2,-\wlane/2) to (-2.5*\wlane,-\wlane/2) to [out=0,in=-90] (\wlane/2,\tvy-\lveh/2-\arlen) to (\wlane/2,\tvy-\lveh/2-\arlen/2);

\draw [thick,draw=blue,fill=blue!20] (\tvx-\lveh/2,\wlane/2-\wveh/2) rectangle ++(\lveh,\wveh)node[pos=.5,] {\small \textbf{TV1}};
\draw[->,thick,draw=blue](\tvx-\lveh/2,\wlane/2) to ++(-\arlen,0);

\draw [thick,draw=blue,fill=blue!20] (\wlane/2-\wveh/2,\tvy-\lveh/2) rectangle ++(\wveh,\lveh)node[pos=.5,rotate=90] {\small \textbf{TV2}};
\draw[->,thick,draw=blue](\wlane/2,\tvy+\lveh/2) to ++(0,\arlen);

\draw [thick,draw=pedestriangreen,fill=pedestriangreen!20] (\pdx-\sped/2,\pdy-\sped/2) rectangle ++(\sped,\sped)node[pos=.5] {\small \textbf{P}};
\draw[->,thick,draw=pedestriangreen](\pdx,\pdy+\sped/2) to ++(0,\arlen);

\end{tikzpicture}
    \caption{Scheme of the urban framework considered in the simulations.}
    \label{fig:urbanroadscheme}
\end{figure}
In this section, we discuss two numerical simulations developed in Matlab and based on the NMPC toolbox~\cite{GrueneEtAl2017}, showing the potential of the proposed control algorithm. We consider an urban intersection, see Fig.~\ref{fig:urbanroadscheme}. Each road has one lane per direction, and the intersection is in the origin of the coordinate system. Units are omitted, as all quantities are given in SI units. The reference path of the EV is the center of the lane, and a B\'ezier curve is used at the intersection. The lane width is $w_\text{lane}=3$, all vehicles have a rectangular shape with $l_\text{veh}=5$ and $w_\text{veh}=2$, and $l_\text{f}=l_\text{r}=2$, and the pedestrian has a squared shape $l_\text{P}=w_\text{P}=1$. The maximum speed allowed is $v_\text{max} = 13$. The input bounds are $\bm{u}_\text{max}=[5,0.52]^\top$, $\bm{u}_\text{min}=[-9,-0.52]^\top$, $\Delta\bm{u}_\text{max}=[9,0.4]^\top$, and $\Delta\bm{u}_\text{min}=-\Delta\bm{u}_\text{max}$. For the TVs, we have $\bm{u}^\text{TV}_\text{max}=[5,0.4]^\top$ and $\bm{u}^\text{TV}_\text{min}=[-9,-0.4]^\top$. We have
\begin{equation}
    \bm{K} =
   \begin{bmatrix}
   0 & k_{12} & 0 & 0\\
   0 & 0 & k_{21} & k_{22}
   \end{bmatrix}
   ,
\end{equation}
with $k_{12}=-0.55$, $k_{21}=-0.63$, and $k_{22}=-1.15$. The high-level feedback gain $\bm{K}_\text{H}$ has the same structure, with $k_{12}^\text{H}=-0.34$, $k_{21}^\text{H}=-0.21$, and $k_{22}^\text{H}=-0.67$. The noise covariance matrices are $\bm{\Sigma}^\text{TV}_{\bm{w}}=\text{diag}(0.15,0.03)$ and $\bm{\Sigma}^\text{P}_{\bm{w}}=\text{diag}(0.05,0.2)$, and $\epsilon_\text{safe}^\text{TV}=4$, $\epsilon_\text{safe}^\text{P}$=1. The risk parameters are $\beta^\text{TV}=0.8$ and $\beta^\text{P}=0.9$ for the trajectory planner, and $\beta_\text{H}^\text{TV}=0.4$ and $\beta_\text{H}^\text{P}=0.5$ for the maneuver planner.

In~\eqref{eqn:LLcostfunction}, $\bm{Q}=\bm{P}=\text{diag}(0,1,1,1)$, $\bm{R}=\text{diag}(0.33,5)$, $\bm{S}=\text{diag}(0.33,15)$, and $r_\text{H}=0.5$ in~\eqref{eqn:HLcostfunction}. The reference speed for the EV is $v_\text{ref}=10$. In all the simulations, $d_\text{ref}=\phi_\text{ref}=0$, the sampling time and prediction horizon are $T=0.2$ and $N=10$ for the trajectory planner, and $T_\text{H}=NT=2$ and $N_\text{H}=8$ for the maneuver planner.

For performance comparison, the following cost function is used, resulting from the computation of the stage cost of~\eqref{eqn:LLcostfunction} for all the $N_\text{sim}$ steps of the simulation:
\begin{equation}
    J_\text{sim}=\sum_{\tau=1}^{N_\text{sim}}\|\Delta\bm{\xi}_\tau\|^2_{\bm{Q}}+\|\bm{u}_\tau\|^2_{\bm{R}}+\|\Delta\bm{u}_\tau\|^2_{\bm{S}}.
    \label{eqn:Jtotsimulation}
\end{equation}
The last component of deviation of the state with respect to the reference trajectory $\Delta\bm{\xi}_{[4]}$ in~\eqref{eqn:Jtotsimulation} is computed considering the desired cruise speed $v_\text{ref}=10$, independently of the optimal reference $\nu_0$ computed by the maneuver planner.

%%%%%%%%%%%%%%%%%%%%%%%%%%%%%%%%%%%%%%%%%%%%%%%%%%%%%%%%%%%%%%%%%%%%%%%%%%%%%%%%
\subsection{Anticipating TV Scenario}
\begin{figure}
\vspace{0.1cm}
    \centering
    \includegraphics[width=0.4\textwidth]{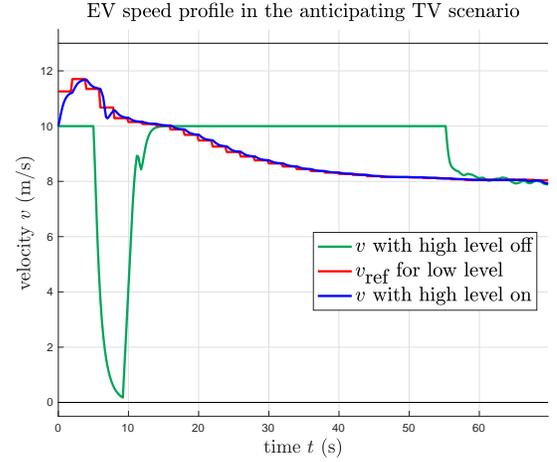}
    \caption{EV speed profile in the anticipating TV scenario.}
    \label{fig:simcarbefore}
\end{figure}
In the first simulation, the EV initial state is $\bm{\xi}_0  = [-70,0,0,10]^\top$, and two TVs are present, one approaching the EV from the right, driving in the opposite direction ($\bm{\xi}_0^\text{TV1}  = [60,-7.5,1.5,0]^\top$), and one driving upward in the vertical road ($\bm{\xi}_0^\text{TV2}  = [1.5,0,10,8]^\top$), see Fig.~\ref{fig:urbanroadscheme}. No pedestrian is considered in this scenario.

If the high-level maneuver planner is not used, the EV drives at cruise speed $v=10$ up to the intersection. Then, TV1 is predicted to occupy the intersection before the EV can safely leave it free, thus the EV stops. Afterwards, the EV turns and proceeds at cruise speed until it reaches TV2, which is driving slower. Eventually, the EV decelerates, in order to keep the safety distance from TV2, based on the risk parameter $\beta^\text{TV}$. By decreasing the risk parameter $\beta^\text{TV}$, the trajectory planner can execute a more aggressive and possibly efficient motion. Conversely, by requiring more safety, the safety distances derived in Section~\ref{subsec:safetydistanceLL} are enlarged, and this ultimately comes at the prices of a higher cost. SMPC permits to balance this trade-off, proving to be notably useful when the disturbances considered are unbounded, and cannot be handled by robust control approaches.

If the maneuver planner is active, the high-level predicts that a stop is necessary at the intersection and acts to prevent it from happening. Hence, a high reference speed is set at first, and the EV reaches the intersection while is still free, before being occupied by TV1. Then, the EV decelerates, reaching the desired cruise speed $v_\text{ref}=10$. Finally, as soon as TV2 is visible, a smooth deceleration is planned.

Fig.~\ref{fig:simcarbefore} shows the speed profile of the EV if the high-level SMPC is inactive (green line) or active (blue), and the optimal reference speed $v_\text{ref}$ (red) dynamically computed by the maneuver planner solving~\eqref{prb:OCPHL}. The maneuver planner allows to optimize the speed, avoiding the stop at the intersection, and making the deceleration smoother when needed. The benefit is also visible in terms of the cost function, since we have $J_\text{sim,1a}=2215.8$ for the first case, and $J_\text{sim,1b}=781.2$ if the maneuver planner is used.

%%%%%%%%%%%%%%%%%%%%%%%%%%%%%%%%%%%%%%%%%%%%%%%%%%%%%%%%%%%%%%%%%%%%%%%%%%%%%%%%
\subsection{Pedestrian Crossing Scenario}

In the second simulation, the EV initial state is $\bm{\xi}_0  = [-100,0,0,10]^\top$, and a pedestrian intends to cross the street in front of the intersection ($\bm{\xi}_0^\text{P}  = [-15,0,-11,1.2]^\top$), see Fig.~\ref{fig:urbanroadscheme}. No TV is considered in this scenario.

Without the maneuver planner, at first the EV proceeds at cruise speed. Then, it rapidly decelerates, to keep the desired distance from the pedestrian, according to the risk parameter $\beta^\text{P}$, until the pedestrian completes crossing. Then, the EV accelerates again to reach the desired cruise speed $v_\text{ref}=10$.

Conversely, if the maneuver planner is used, the EV stop caused by the crossing pedestrian is predicted. Yet, now the maneuver planner cannot set the speed in order to safely pass before the pedestrian crosses, and therefore an initial deceleration is planned. In such a way, the EV can safely pass the crossing point without stopping, since the pedestrian is already on the other side of the street.

\begin{figure}
\vspace{0.1cm}
    \centering
    \includegraphics[width=0.4\textwidth]{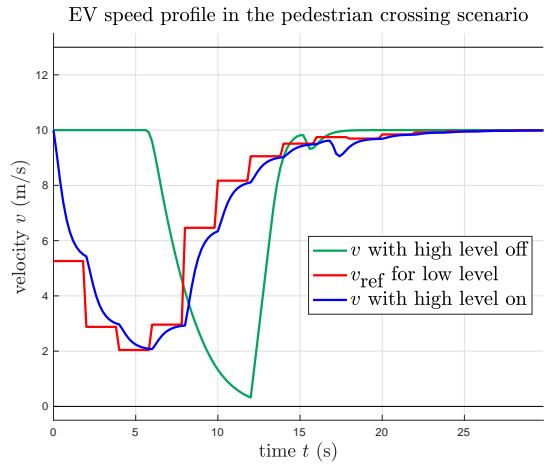}
    \caption{EV speed profile in the pedestrian crossing scenario.}
    \label{fig:simpedestrian}
\end{figure}

The speed profile of the EV for this scenario is shown in Fig.~\ref{fig:simpedestrian}, if the maneuver planner is inactive (green line) or active (blue). The optimal reference speed $v_\text{ref}$ is represented in red. Even though the maneuver planner effectively avoids the full stop, the speed is significantly decreased in the first stretch, and the overall cost is $J_\text{sim,2b}=2049.2$, which is slightly higher than in the cost in the first case, $J_\text{sim,2a}=1992.0$, if the high-level maneuver planner is inactive.

%%%%%%%%%%%%%%%%%%%%%%%%%%%%%%%%%%%%%%%%%%%%%%%%%%%%%%%%%%%%%%%%%%%%%%%%%%%%%%%%
\section{CONCLUSION}
\label{sec:conclusion}
In this work we applied SMPC to urban autonomous driving. A proper choice of the risk parameter $\beta$ allows to balance between safety and conservativeness. In the trade-off, requiring the safety constraints hold only up to a certain probability allows to plan an efficient motion, whereas the risk level is kept limited. Moreover, the hierarchical control structure is beneficial since it allows to plan separately the short-term trajectory and the long-term maneuver (i.e., the speed in the considered example). As a result, the speed of the autonomous vehicle can be set so as to smooth the motion, taking into account foreseen future stops caused by other road participants, and to avoid stops if possible.

Concerning future research, this method could be applied to more complex scenarios, where a richer set of high-level maneuvers must be considered, e.g., lane change decisions. Moreover,  formal conditions guaranteeing that this two-level optimization framework is effective should be investigated.

%%%%%%%%%%%%%%%%%%%%%%%%%%%%%%%%%%%%%%%%%%%%%%%%%%%%%%%%%%%%%%%%%%%%%%%%%%%%%%%%
%\addtolength{\textheight}{-3.6cm}   % This command serves to balance the column lengths

%%%%%%%%%%%%%%%%
%%%%%%%%%%%%%%%%
\bibliographystyle{IEEEtran}
\bibliography{res}

\begin{thebibliography}{10}
\providecommand{\url}[1]{#1}
\csname url@samestyle\endcsname
\providecommand{\newblock}{\relax}
\providecommand{\bibinfo}[2]{#2}
\providecommand{\BIBentrySTDinterwordspacing}{\spaceskip=0pt\relax}
\providecommand{\BIBentryALTinterwordstretchfactor}{4}
\providecommand{\BIBentryALTinterwordspacing}{\spaceskip=\fontdimen2\font plus
\BIBentryALTinterwordstretchfactor\fontdimen3\font minus
  \fontdimen4\font\relax}
\providecommand{\BIBforeignlanguage}[2]{{%
\expandafter\ifx\csname l@#1\endcsname\relax
\typeout{** WARNING: IEEEtran.bst: No hyphenation pattern has been}%
\typeout{** loaded for the language `#1'. Using the pattern for}%
\typeout{** the default language instead.}%
\else
\language=\csname l@#1\endcsname
\fi
#2}}
\providecommand{\BIBdecl}{\relax}
\BIBdecl

\bibitem{ChenEtAl2019b}
J.~Chen, B.~Yuan, and M.~Tomizuka, ``Model-free deep reinforcement learning for
  urban autonomous driving,'' in \emph{IEEE 22nd Intelligent Transportation
  Systems Conference (ITSC)}, 2019.

\bibitem{HawkeEtAl2020}
J.~{Hawke}, R.~{Shen}, C.~{Gurau}, S.~{Sharma}, D.~{Reda}, N.~{Nikolov},
  P.~{Mazur}, S.~{Micklethwaite}, N.~{Griffiths}, A.~{Shah}, and A.~{Kndall},
  ``Urban driving with conditional imitation learning,'' in \emph{IEEE
  International Conference on Robotics and Automation (ICRA)}, 2020.

\bibitem{ChenEtAl2019}
J.~{Chen}, B.~{Yuan}, and M.~{Tomizuka}, ``Deep imitation learning for
  autonomous driving in generic urban scenarios with enhanced safety,'' in
  \emph{IEEE/RSJ International Conference on Intelligent Robots and Systems
  (IROS)}, 2019.

\bibitem{LevinsonEtAl2011}
J.~{Levinson}, J.~{Askeland}, J.~{Becker}, J.~{Dolson}, D.~{Held}, S.~{Kammel},
  J.~Z. {Kolter}, D.~{Langer}, O.~{Pink}, V.~{Pratt}, M.~{Sokolsky},
  G.~{Stanek}, D.~{Stavens}, A.~{Teichman}, M.~{Werling}, and S.~{Thrun},
  ``Towards fully autonomous driving: Systems and algorithms,'' in \emph{IEEE
  Intelligent Vehicles Symposium (IV)}, 2011.

\bibitem{SchildbachEtAl2016}
G.~{Schildbach}, M.~{Soppert}, and F.~{Borrelli}, ``A collision avoidance
  system at intersections using robust model predictive control,'' in
  \emph{IEEE Intelligent Vehicles Symposium (IV)}, 2016.

\bibitem{BruedigamEtAl2021}
T.~Br\"udigam, M.~Olbrich, D.~Wollherr, and M.~Leibold, ``Stochastic model
  predictive control with a safety guarantee for automated driving,''
  \emph{IEEE Transactions on Intelligent Vehicles}, 2021.

\bibitem{NguyenEtAl2017}
N.~A. {Nguyen}, D.~{Moser}, P.~{Schrangl}, L.~{del Re}, and S.~{Jones},
  ``Autonomous overtaking using stochastic model predictive control,'' in
  \emph{11th Asian Control Conference (ASCC)}, 2017.

\bibitem{CarvalhoEtAl2014}
A.~Carvalho, Y.~Gao, S.~Lefevre, and F.~Borrelli, ``Stochastic predictive
  control of autonomous vehicles in uncertain environments,'' in \emph{12th
  InternationaI Symposium on Advanced Vehicle Control}, 2014.

\bibitem{SuhEtAl2018}
J.~{Suh}, H.~{Chae}, and K.~{Yi}, ``Stochastic model-predictive control for
  lane change decision of automated driving vehicles,'' \emph{IEEE Transactions
  on Vehicular Technology}, 2018.

\bibitem{BaethgeEtAl2016}
T.~{Bäthge}, S.~{Lucia}, and R.~{Findeisen}, ``Exploiting models of different
  granularity in robust predictive control,'' in \emph{IEEE 55th Conference on
  Decision and Control (CDC)}, 2016.

\bibitem{BruedigamEtAl2020b}
T.~Brüdigam, J.~Teutsch, D.~Wollherr, and M.~Leibold, ``Combined robust and
  stochastic model predictive control for models of different granularity,'' in
  \emph{21th IFAC World Congress}, 2020.

\bibitem{ShekharEtAl2015}
R.~C. Shekhar and C.~Manzie, ``Optimal move blocking strategies for model
  predictive control,'' \emph{Automatica}, 2015.

\bibitem{El-FekyEtAl2017}
S.~M. {El-Feky}, A.~M. {Zaki}, A.~M. {Bahaa-Eldin}, and M.~H. {El-Shafey},
  ``Improving model predictive controller by using non-uniform sampled model,''
  in \emph{International Conference on Advances in Computing, Communications
  and Informatics (ICACCI)}, 2017.

\bibitem{BrdysEtAl2008}
M.~Brdys, M.~Grochowski, T.~Gminski, K.~Konarczak, and M.~Drewa, ``Hierarchical
  predictive control of integrated wastewater treatment systems,''
  \emph{Control Engineering Practice}, 2008.

\bibitem{DinhEtAl2020}
N.~Dinh, M.~Sualeh, D.~Kim, and G.-W. Kim, ``A hierarchical control system for
  autonomous driving towards urban challenges,'' \emph{Applied Sciences}, 2020.

\bibitem{ZieglerEtAl2014}
J.~{Ziegler}, P.~{Bender}, M.~{Schreiber}, H.~{Lategahn}, T.~{Strauss},
  C.~{Stiller}, T.~{Dang}, U.~{Franke}, N.~{Appenrodt}, C.~G. {Keller},
  E.~{Kaus}, R.~G. {Herrtwich}, C.~{Rabe}, D.~{Pfeiffer}, F.~{Lindner},
  F.~{Stein}, F.~{Erbs}, M.~{Enzweiler}, C.~{Knöppel}, J.~{Hipp}, M.~{Haueis},
  M.~{Trepte}, C.~{Brenk}, A.~{Tamke}, M.~{Ghanaat}, M.~{Braun}, A.~{Joos},
  H.~{Fritz}, H.~{Mock}, M.~{Hein}, and E.~{Zeeb}, ``Making bertha drive—an
  autonomous journey on a historic route,'' \emph{IEEE Intelligent
  Transportation Systems Magazine}, 2014.

\bibitem{WangEtAl2015}
Q.~Wang, T.~Weiskircher, and B.~Ayalew, ``Hierarchical hybrid predictive
  control of an autonomous road vehicle,'' in \emph{ASME Dynamic System and
  Control Conference}, 2015.

\bibitem{CurvedBicycleDynamics}
M.~{Werling} and L.~{Groll}, ``Low-level controllers realizing high-level
  decisions in an autonomous vehicle,'' in \emph{IEEE Intelligent Vehicles
  Symposium}, 2008.

\bibitem{MizushimaEtAl2019}
Y.~Mizushima, I.~Okawa, and K.~Nonaka, ``Model predictive control for
  autonomous vehicles with speed profile shaping,'' in \emph{10th IFAC
  Symposium on Intelligent Autonomous Vehicles IAV}, 2019.

\bibitem{MuraleedharanEtAl2020}
A.~{Muraleedharan}, A.~T. {Tran}, H.~{Okuda}, and T.~{Suzuki}, ``Scenario-based
  model predictive speed controller considering probabilistic constraint for
  driving scene with pedestrian,'' in \emph{IEEE 23rd International Conference
  on Intelligent Transportation Systems (ITSC)}, 2020.

\bibitem{GrueneEtAl2017}
L.~Gr\"une and J.~Pannek, \emph{Nonlinear Model Predictive Control}.\hskip 1em
  plus 0.5em minus 0.4em\relax Springer, 2017.

\end{thebibliography}

\end{document}